\documentstyle[12pt]{article}
\input epsf.tex

\newcommand{\be}{\begin{equation}}
\newcommand{\ee}{\end{equation}}
\newcommand{\ba}{\begin{eqnarray}}
\newcommand{\ea}{\end{eqnarray}}
\newcommand{\pa}{\partial} 

\def\bea{\begin{eqnarray}}
\def\eea{\end{eqnarray}}

\def\bb#1{\hbox{\mybb#1}}

\def\zet{\bb{Z}}
\def\real{\bb{R}}

\def\R4{\bb{R}^4}
\font\mybb=msbm10 at 12pt

\def\C{{\cal C}}
\def\Hol{{\mbox{Hol}}}
\def\tr{{\mbox{Tr}}}
\def\Link{{\mbox{Link}}}
\def\sLink{{\mbox{sLink}}}

\def\unita{{1 \kern-.30em 1}}

\begin{document}

\begin{titlepage}
\begin{flushright}
{IFUM 526/FT}\\
{ROM2F/96/22}\\
\end{flushright}
\vskip 1mm
\begin{center}
 
{\large \bf  The BF Formalism  for QCD  and Quark Confinement}\\ 
  
\vspace{0.5cm}
{\bf Francesco Fucito}
\vskip 0.1cm
\centerline{\sl Dipartimento di Fisica, Universit\`a di Roma II ``Tor Vergata"}
\centerline{\sl and}  
\centerline{\sl I.N.F.N. \ - \  Sezione di Roma II, }
\centerline{\sl Via Della Ricerca Scientifica \ \ 00133 \ Roma \ \ ITALY}
\vskip .2cm
\centerline{{\bf Maurizio Martellini}
\footnote{On leave of absence from Dipartimento di Fisica, Universit\`a 
di Milano, 20133 Milano, ITALY. 
E-mail: {\it martellini@vaxmi.mi.infn.it, martellini@nbivax.nbi.dk}}}
\vskip 0.1cm
\centerline{\sl The Niels Bohr Institute, University of Copenhagen}
\centerline{\sl DK-2100 Copenhagen $\phi$, Denmark, }
\centerline{\sl I.N.F.N.\ - \ Sezioni di Pavia e Milano} 
\centerline{\sl and}
\centerline{\sl Landau Network  at ``Centro Volta'', Como, ITALY}
\vskip .2cm
 {\bf Mauro Zeni}
\footnote{E-mail:{\it zeni@vaxmi.mi.infn.it}}
 
\vspace{0.1cm}

{\sl Dipartimento di Fisica, Universit\`a di Milano \\
and \\  
I.N.F.N. \ - \  Sezione di Milano, \\
Via Celoria 16 \ \ 20133 \ Milano \ \ ITALY}
\end{center}
\abstract{
Using the BF version of pure Yang-Mills, it is possible to find  a covariant
representation of the 't Hooft magnetic flux operator. In this framework, 
't Hooft's pioneering work on confinement finds an explicit realization in the
continuum. Employing the Abelian projection gauge we compute the expectation
value of the magnetic variable and find the expected perimeter law. 
We also check the
area law behaviour for the Wilson loop average and compute the string 
tension which 
turns out to be of the right order of magnitude.}
\vfill
\end{titlepage}
\setcounter{section}{0}
\section{Introduction}
\setcounter{equation}{0}
\addtolength{\baselineskip}{0.3\baselineskip} 
The problem of quark confinement in continuum QCD is still unsolved. 
Confinement of electric charges is observed if one is willing to use 
the lattice as 
a regulator but
direct computations in the continuum theory are not easily performed due to the
large scale nature of the problem at hand.

Among the different explanations put forward in literature, we find
that the so called dual Meissner effect is the most appealing and complete:
if the QCD vacuum behaves as a magnetic superconductor color electric 
flux lines
will form long and narrow tubes with color electric charges at its ends.
The story of this idea is quite long and has received contributions from many
authors \cite{no,mandel,thsc,poly}. In this paper we will stick to 
the version of 't Hooft 
in which all
these ideas are brought to a high degree of definitiveness 
\cite{thooft1,thooft2}.

More in detail, the whole idea of  the dual superconductor can be 
phrased in terms
of 't Hooft order-disorder variables. These variables are non-local quantum 
operators, $W(C), M(C^\prime)$, creating 
thin color electric and magnetic fluxes, along the curves $C, C^\prime$, 
in the QCD vacuum. 
In their turn, the curves $C, C^\prime$ border two surfaces $\Sigma, 
\Sigma^\prime$.
In the absence of massless particles, these surfaces may become dynamical 
and we may
have linkage when one of these curves (say $C^{\prime}$ for example) 
intersects the surface 
(say $\Sigma$)  spanned by the  other curve ($C$). The 
linkage costs
energy per unity of area and, roughly speaking,  the associated 
effective action
goes as the area, $A$, of the dynamical surface: in our 
previous
example it will be $<W(C)>\simeq e^{const.A(\Sigma)}$.
All these properties are
summarized by the commutation relation of  $W(C), M(C^\prime)$
which is of braiding 
type and which
contains the entire physics of the problem. Which surface is intersecting 
which curve
is not known in this picture which supports all possibilities: Higgs phase, 
Coulomb phase,
partial Higgs with confinement and confinement phase \cite{thooft1}. 
To support the picture of the dual superconductor the QCD vacuum must 
be filled 
by magnetic vortex lines with non-trivial topology, generated by means of 
a true or effective dual Higgs dynamics. 

Despite the appeal of this qualitative picture one may still be left  with 
a feeling of uneasiness since quantitative computation in the continuum 
theory have not yet been performed.
 
In order to expose 
the 
topological
content of the theory, one of us has proposed to study pure Yang-Mills 
theory using
the first order formalism (BF-YM). The main advantage of this 
formulation
is that a color magnetic  operator can be easily guessed in the continuum  
\cite{ccgm}. In this formalism we show that our magnetic operator is the dual 
't Hooft observable \cite{thooft1} and compute its $vev$, employing the Abelian 
projection gauge \cite{thooft2,eza}, finding the expected perimeter law.

The first order form of pure Yang-Mills is described by the action 
functional
\be 
S_{BF-YM} = \int \tr (iB\wedge F +{g^2\over 4}B\wedge *B) 
\label{uno.1}
\ee 
where 
\be
F\equiv {1\over 2}F^a_{\mu\nu}T^a dx^\mu\wedge dx^\nu\equiv dA+
i[A,A]\quad ,
\label{uno.1bis}
\ee  
$D\equiv d +i[A,\cdot ]$ and $B$ is a Lie valued 2-form. 
The generators of the $SU(N)$ Lie 
algebra in the fundamental representation are normalized as 
$\tr T^aT^b=\delta_{ab}/2$ and the $*$ product (Hodge duality)
for a $p$ form in $d$ dimension is defined as 
$*=\epsilon^{i_1\ldots i_d}/(d-p)!$. 
The classical gauge invariance of (\ref{uno.1}) is given by 
\ba 
\delta A &=& D\Lambda_0\quad , \nonumber \\ 
\delta B &=& -i[\Lambda_0 ,B]\quad . 
\label{uno.5} 
\ea 

Using the field equations 
(which is  equivalent in the path integral to integrate over B)
\ba 
 F& =&{ig^2\over 2}*B\quad ,
\label{uno.2}\\ 
DB&=&0\quad , \nonumber 
\ea  
we get the standard YM action
\be  
S_{YM} ={1\over g^2} \int \tr (F\wedge *F) \quad .
\label{uno.3}
\ee  
Note that the relation between the formulations (\ref{uno.1})
and (\ref{uno.3}) may be also undertood as a duality map; however 
off shell  $B$ does not satisfy a Bianchi identity and this fact will 
be connected with 
the introduction in the theory of magnetic vortex lines.
We remark that the short distance quantum behaviour of (\ref{uno.1})
and (\ref{uno.3}) are the same as it has been explicitely checked 
\cite{mz}.

We remark that the action functional in (\ref{uno.1}) defined by
\be  
S_{BF}=i\int \tr (B\wedge F)
\label{uno.4} 
\ee 
is known as the 4D pure bosonic BF-theory and defines a 
topological quantum field theory \cite{horouno,blau}. 
The form of (\ref{uno.1}) was understood in \cite{ccgm} as an 
indication that 
the bosonic YM theory can be viewed as a {\it perturbative expansion} 
in the  
coupling $g$ around the topological pure BF theory (\ref{uno.4}). 
This 
procedure has been called topological embedding \cite{ans}.
Using this perturbative picture around BF-theory we check 
in this paper the area law behaviour 
for the Wilson operator.

The plan of the paper is the following: in section 2 we introduce an
explicit analytic realization of the 't Hooft magnetic variable in terms 
of the $A, B$ fields and check its properties at the operator level using
the canonical quantization. In section 3 we compute its expectation value
in the abelian projection gauge and verify the perimeter law. In section
4 we verify the area law for the Wilson loop operator. At last, in section
5 we draw our conclusions.
\section{A Color Magnetic Operator}
\setcounter{equation}{0}
In this section we define a  gauge invariant non local  
operator $M(C)\equiv M(\Sigma ,C)$ associated with a fixed orientable surface 
$\Sigma$ in $M^4$ (our base manifold) and with a suitable choice of a closed 
contour $C$ on $\Sigma$, which gives an explicit realization of 
the `t Hooft loop.
Part of this section
was already investigated in \cite{ccgm}  by one of the present authors.  

The presence of the Lie-algebra valued two-form $B$ field in (\ref{uno.1})
allows the definition of the observable gauge invariant operator
\be                       
M(\Sigma ,C)\equiv \tr \exp \{ik
\int_{\Sigma } d^2y 
\ \Hol^y_{\bar x} (\gamma ) B(y) \Hol_y^{\bar x} (\gamma^{\prime} )\} \quad ,
 \label{tre.2} 
\ee 
where $k$ is an arbitrary expansion parameter, $\bar x$ is a {\it fixed} 
point  over the orientable surface $\Sigma\in M^4$ (we do not 
integrate over $\bar x$) and the relation between the assigned paths 
$\gamma$, $\gamma^{\prime}$ over $\Sigma$ and the closed contour $C$ is the 
following: let $\hat \Sigma$ to be a piecewise
linear (PL) approximation of 
$\Sigma$ by 
plaquettes (any two dimensional topological variety admits a 
PL decomposition). Then the closed path $C$ is given by a 
succession of subpaths  formed 
by the 1-skeleton of the 2-PL manifold $\hat\Sigma$. In other words $C$ 
starts from the fixed point $\bar x$, connects a point $y\in\Sigma$ by 
the open path $\gamma_{\bar x y}$ and then returns back to the
neighborhood of $\bar x$ by 
$\gamma_{y\bar x }^{\prime}$, 
(which is not restricted to coincide with the inverse 
$(\gamma_{\bar x y})^{-1}=\gamma_{y\bar x }$). From the neighborhood 
of $\bar x$ the path starts again to connect another point 
$y^{\prime}\in\Sigma$. Then it 
returns back to the neighborhood of $\bar x$ and so on until all points
on $\Sigma$ are connected. The path 
$C_{\bar x}=\{ \gamma\cup\gamma^{\prime}\} $ is generic and we do not require 
any particular ordering prescription as it is done in similar constructions 
devoted to obtain a non abelian Stokes theorem \cite{ara}.
In figure 1 we show a typical path $C$ connecting four points $x1, x2, x3, x4$
on a surface $\Sigma$ given by a plane.
\vskip .6cm
\centerline{\vbox{\epsfysize=40mm \epsfbox{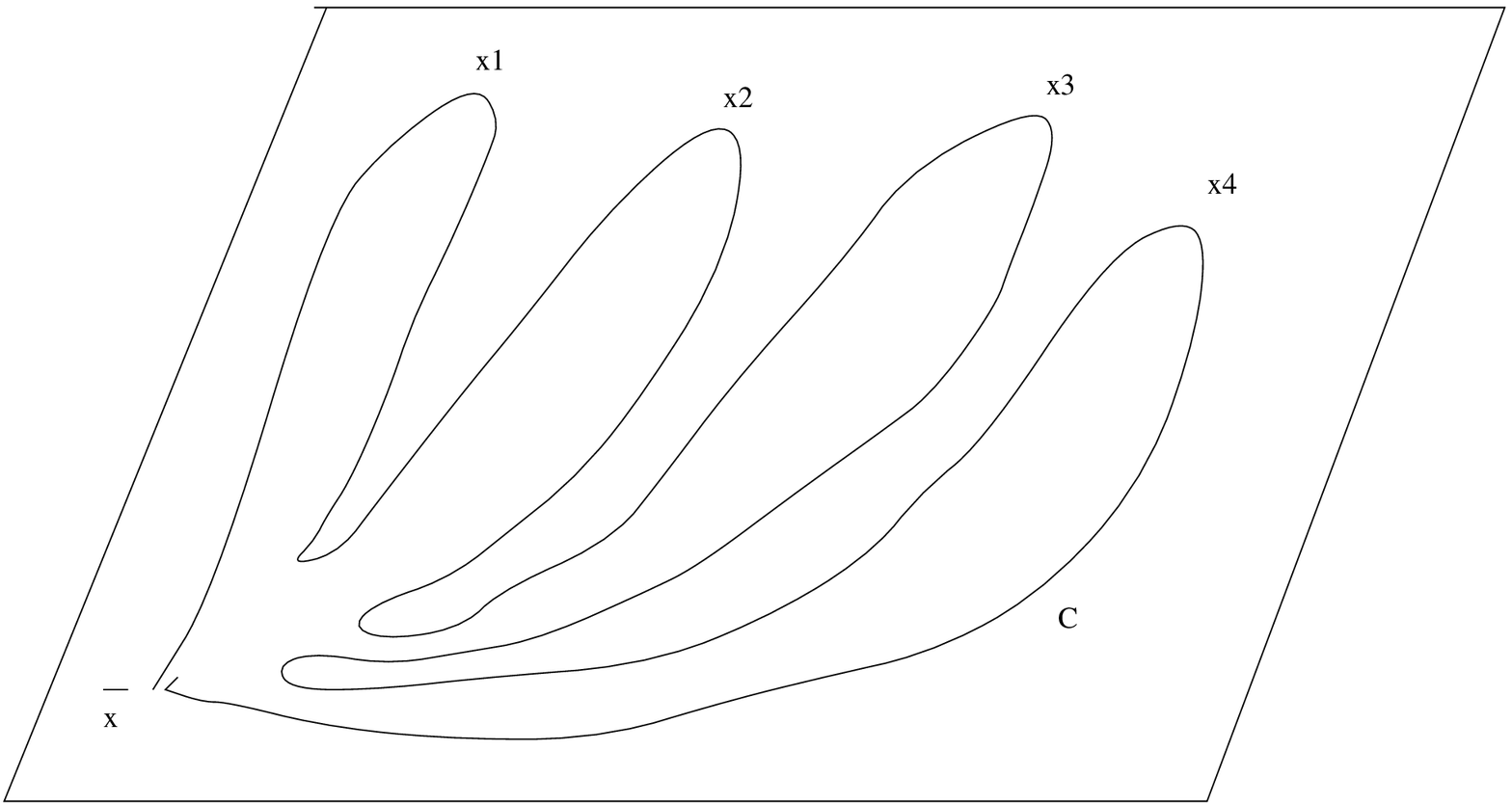}}}
\smallskip
\vskip .4cm
\centerline{\bf Figure 1}
\vskip .6cm
Of course the 
quantity (\ref{tre.2}) is path dependent and our strategy is to regard it as 
a loop variable once the surface $\Sigma$ is given. 
In Eq. (\ref{tre.2}) $\Hol_{\bar x}^y(\gamma )$ denotes the usual holonomy 
along the open path $\gamma\equiv\gamma_{\bar x y}$ with initial and final 
point $\bar x$ and $y$ respectively,
\be 
\Hol_{\bar x}^y(\gamma )\equiv P\exp (i\int_{\bar x}^y dx^{\mu} A_{\mu}(x))
\quad .
\label{tre.3}
\ee 
Given the finite local gauge transformations 
\ba 
\Hol_{\bar x}^y &\rightarrow & g^{-1}(\bar x) \Hol_{\bar x}^y g(y) \quad , 
\label{tre.4} \\ 
B(y)  &\rightarrow & g^{-1}(y) B(y) g(y) \quad  ,
\label{tre.5} 
\ea 
it is immediate to prove the gauge invariance of (\ref{tre.2}). 
A comment on the geometrical meaning of the operator $M$ is in order. 
If, for example,  we take the surface $\Sigma$ to be a torus we may 
define $M(\Sigma,C)$ in terms of the
parallel transport operator of loops \cite{poly} given by 
$P \exp \{ik\oint_{C_1} 
\tilde {\cal B}_{\mu} (C_1)dx^{\mu}\}$, 
where $\tilde{\cal B}_{\mu}(C_1)\equiv \oint_{\C_2} (Hol B_{\mu\nu} 
Hol^{\prime})dx^{\nu}$ is a connection in the loop space and $C$ is a 
linear combination of the two fundamental cycles $C_1$ and $C_2$ 
of the torus.

By adding suitable bosonic vector and 
ghosts fields to the 
BF-YM theory, the ``topological" (we will clarify later 
what we use quotes here) symmetry of the 
pure BF action (\ref{uno.4}),
given by
\ba 
\delta A(y) &=& 0 \quad ,\label{tre.6} \\ 
\delta B(y) &=& D\Lambda_1 (y)
\equiv d_{(y)}\Lambda_1(y) +i[A(y),\Lambda_1(y)]\quad ,
\nonumber
\ea 
could be extended to the BF-YM theory. Then by using the well-known 
identity \cite{gasio}
\be 
d_{(y)} (\Hol_{\bar x}^y (\gamma )\Lambda_1(y)
\Hol^{\bar x}_y(\gamma^{\prime})) 
=\Hol_{\bar x}^y (\gamma )\{ d_{(y)}\Lambda_1(y)+i[A(y),\Lambda_1(y)]\} 
\Hol^{\bar x}_y (\gamma^{\prime})
\label{tre.7} 
\ee 
one may prove the invariance of 
(\ref{tre.2}) under the transformations (\ref{tre.6}) up to boundary 
terms for non-compact $\Sigma$. 
However these boundary terms may be neglected  in the 
renormalized theory if mixed Dirichlet-Neumann boundary 
conditions for the quantum fields at the boundary 
$\pa (M^4 \backslash\Sigma )=-\pa\Sigma $ are used, 
in a way similar to the quantization of systems over a non complete 
space time (e.g. Casimir effect). 

Let us give a geometrical interpretation of the 
``topological" symmetry (\ref{tre.6}). The holonomy dressed field 
$\Hol B \Hol^{\prime}$ can be interpreted as a colored charged field 
that 
transforms covariantly under the global  colour symmetry $g(\bar x)$. 
Accordingly the transformation (\ref{tre.6}) represents a local 
change (with parameter $\Hol \Lambda_1 \Hol^{\prime}$) 
of the colour charge \cite{faridani}.

In the following the path dependence of (\ref{tre.2}) 
from the closed contour $C$ will be crucial. Indeed, we shall see below that 
different choices of $C$ for fixed surface 
$\Sigma$,  we call it a framing $(C,\Sigma )$,
 give different color magnetic fluxes in QCD.

The classical BF-YM 
action (\ref{uno.1}) is linear in time derivatives, and 
hence it is easily cast into its canonical form \cite{horouno}. 
Let $M^4 =N^3\times\real$ and let $t$ be a coordinate labelling
the different $N^3$ surfaces. Let then $\xi^\mu$ be any vector satisfying
$\xi^\mu\partial_\mu t=1$. The appropriate notion of time derivative
is then given  by the Lie derivative
\be
L_\xi A^a_\mu =\partial_\mu(A^a_\nu\xi^\mu)+\xi^\nu\partial_\nu A^a_\mu-
\xi^\nu\partial_\mu A^a_\nu=D_\mu(A^a_\nu\xi^\nu)+\xi^\nu F_{\nu\mu}
\label{tre.7bis}\quad .
\ee 
For $N^3=\real^3$ and $\xi_\mu=(1,0,0,0)$ one recovers the usual definition
of time derivative.
Time derivatives act only on the 
space components of $A_{\mu}^a$. Therefore the conjugate momenta 
of $A^a_0, B_{0i}$
are zero and these fields appear as 
Lagrangian multipliers since they may be eliminated through the classical 
constraint equations. 
In order to have first class constraints in the Hamiltonian formalism,
the field content of the action (\ref{uno.1}) must be enlarged in such
a way that the resulting action enjoys the same topological symmetry
(\ref{tre.6}) of the action (\ref{uno.4}). We thus introduce a Lie valued
vector field $\eta$ which transforms under the symmetry (\ref{tre.6})
as $\delta\eta=\Lambda_1$ and enters the action (\ref{uno.1}) by
substituting
$B^2\rightarrow (B-D\eta)^2$ \cite{noi}.

The 3+1-action becomes
\ba 
S_{BF-YM}&=&{1\over 2}\int dt\int_{N^3} d^3x
\{2i\epsilon^{ijk}B^a_{ij}\dot A^a_k
+{g^2\over 2}[B^a_{0i}-(D_0\eta^a_i-D_i\eta^a_0)]^2
\nonumber\\
&-&i\epsilon^{ikl}B^a_{0i}F^a_{kl} 
+iA_0^a \epsilon^{ijk}(D_iB_{jk})^a
+{g^2\over 4}[B^a_{ij}-(D_i\eta^a_j-D_j\eta^a_i)]^2\} 
\label{tre.10}
\ea 
and the classical constraints read 
\ba 
{\cal C}_1[\tau (\vec x)] &\equiv & {1\over 2}
\int_{N^3}d^3x \ i\tau^a\epsilon^{ijk} (D_{i}B_{jk})^a=0\quad ,
\label{tre.11} 
\\ 
{\cal C}_2 [v_i(\vec x)] &\equiv & {1\over 2}
\int_{N^3}d^3x \ v^a_i 
(\Pi^a_i-i\epsilon^{ikl}F^a_{kl})=0 \quad ,
\label{tre.12}
\ea 
where $a$ is the Lie algebra index, $i,j,k=1,2,3$ and  
$\tau^a ,v^a_l$ are arbitrary  Lie algebra valued fields on $N^3$. 
$\Pi^a_i$ is the canonical momentum conjugate to $\eta$
\be
\Pi^a_i={g^2\over 2}\{B^a_{0i}-\partial_0\eta^a_i+i[A_0,\eta_i]^a-i
[A_i,\eta_0]^a\}\quad ,
\label{tre.12ter}
\ee
and
\be
\chi^a_i=i\epsilon_{ijk}B_{jk}^a
\label{tre.12bis}
\ee
is the canonical momentum conjugate to $A^a_i$.

The appropriate gauge for the Hamiltonian formalism is 
the temporal gauge,
which in this contest reads
\ba 
A_0^a &=& 0 \quad , \label{tre.13} \\
B_{0i}^a &=& 0 \quad .  
\label{tre.14} 
\ea 
(\ref{tre.13}) fixes the conventional gauge symmetry
and (\ref{tre.14}) fixes the topological symmetry.

The Poisson brackets are defined by 
\ba
\{ A^a_i (\vec x),\chi^b_j (\vec y)\}_{x_0=y_0} =
i\delta^{ab}\delta_{ij}
\delta^{(3)}(\vec x -\vec y),\nonumber\\
\{\eta_i^a(\vec x),\Pi^b_j(\vec y)\}=
\delta^{ab}\delta_{ij}\delta^{(3)}(\vec x -\vec y),
\label{tre.16} 
\ea
and yield 
\ba 
\delta  A^a_i (\vec x) &&=\{{\cal C}_1, A^a_i (\vec x)\} +
\{{\cal C}_2, A^a_i (\vec x)\} =-2iD_i\tau^a (\vec x) \quad ,
\nonumber\\  
\delta  B^a_{ij} (\vec x) &&=\{{\cal C}_1, B^a_{ij} (\vec x)\} +
\{{\cal C}_2, B^a_{ij} (\vec x)\} = 
-2i[\tau (\vec x),B_{ij}(\vec x)]^a \nonumber\\
&&-2i(D_{i}v^a_{j} (\vec x)-
D_{j}v^a_{i} (\vec x))\quad ,\label{tre.17}\\
\delta\eta_i^a &&^=
\{{\cal C}_1, \eta^a_i(\vec x)\} +
\{{\cal C}_2, \eta^a_i (\vec x)\} = 
-i[\tau (\vec x),\eta_i(\vec x)]^a-v^a_i(\vec x)\quad .
\nonumber 
\ea 
The constraint ${\cal C}_1$ generates the usual non 
abelian gauge 
transformations for $A^a_i$, $\eta^a_i$ and $B^a_{ij}$, 
and ${\cal C}_2$ generates 
the classical topological symmetry (\ref{tre.6}). 
The canonical quantization of  (\ref{tre.10}) is 
obtained by replacing 
the Poisson bracket $\{\cdot ,\cdot \}$ in 
(\ref{tre.16}) with 
the commutator $-i\hbar [\cdot ,\cdot ]$ and 
promoting the classical 
constraints ${\cal C}_1,{\cal C}_2$ to 
operator valued ones
\ba 
\hat {\cal C}_1 |A,B,\eta>&=0 
\quad ,\label{tre.18} \\ 
\hat {\cal C}_2 |A,B,\eta>&=0 \quad ,
\nonumber 
\ea 
where $|A,B,\eta>$ is 
called  a physical state if it satisfies the above quantum 
constraints.
We now show that
$M(C)\equiv M(C,\Sigma =S^2)$  generates a local singular (or
equivalently a multivalued
regular) gauge transformation, $\Omega_C (\vec x)$, along $C$: this 
is precisely  
the defining
property of the `t Hooft  color magnetic variable \cite{thooft1}. 

We start from the classical quantity associated to (\ref{tre.2}) 
and assume 
that $\Sigma $, and hence $C$, is small, so that one may consider 
a Taylor 
expansion of (\ref{tre.2}) around $k=0$. By using the Gauss 
theorem, the gauge (\ref{tre.13}), (\ref{tre.14}) and the 
identity (\ref{tre.7}), which in our case becomes
\ba 
&& \tr [\int_{S^2} d\sigma^{ij}(\Hol B_{ij}\Hol^{\prime})] =
\tr [\int_{M^3} d^3x\epsilon^{ijk}\pa_{i} 
(\Hol B_{jk}\Hol^{\prime})]= \nonumber \\ 
&& =\tr [\int_{M^3} d^3x(\Hol\  \epsilon^{ijk} 
D_{i} B_{jk}\Hol^{\prime})]=
\tr [\int_{M^3} d^3x\Hol^{-1}(C_x) \epsilon^{ijk} 
D_{i} B_{jk}]\quad ,
\label{tre.18bis} 
\ea 
one gets  at lowest order in $k$ 
\be 
M(C)\simeq \tr \{ \unita +ik\int_{N^3}d^3x 
[\Hol^{-1}(C_x) \epsilon^{ijk} D_{i} B_{jk}(x)] \} 
\quad ,\label{tre.19}
\ee 
where
 $M^3$ is a small 3-ball with boundary $S^2:\pa M^3 =S^2$ and 
$C_x$ is defined as $C_x\equiv \gamma_x^{\bar x}\cup
\gamma_{\bar x}^{\prime x}$. 
We now expand also  $\Hol^{-1}$ at this 
order
\be 
\Hol^{-1}(C_x)\simeq (\unita -i\varphi_C (x))\qquad ,
\qquad \varphi_C(x)\equiv\oint_{C_x}dy^{i}A_{i}\qquad .
\label{tre.20}
\ee 
Substituting (\ref{tre.20}) in (\ref{tre.19}) and using the definition of the 
constraint ${\cal C}_1$, Eq. (\ref{tre.11}), we see that $M(C)$ 
becomes the generator 
of the classical infinitesimal local gauge transformations with parameter    
$k\varphi_C(x)$:
\be 
M(C)\simeq \tr \{ \unita+k\tilde {\cal C}_1 [\varphi_C(\vec x)] \} \quad ,
\label{tre.21}
\ee 
where ${\cal C}_1=2\tr  [\tilde {\cal C}_1]$ in the temporal gauge
\footnote{
Notice that $M(C)$ is the
non-abelian generalization of the observable  carrying 
non trivial magnetic charge studied  in Ref.\cite{mint,digia}.
Indeed the variable $\pi^i$ defined in eq.(14) of Ref.\cite{mint} and in
eq.(10) of Ref.\cite{digia} is the conjugate momentum of the 
vector potential which, in our first order formalism, is to be identified
with the $B$ two-form. Their monopole magnetic charge
becomes, in our formalism, $\phi_C( x)$. See also next chapter.
}.

To extend (\ref{tre.21}) to the quantum level we substitute
the Poisson brackets with the quantum mechanical commutators and 
rescale the geometrical fields to the physical ones, 
\ba 
&& \hat A^{a}_i(\vec y)=g\hat A^{\prime a}_i(\vec y)\quad ,
\label{tre.22} \\ 
&& [\hat A^{\prime a}_i(\vec y),\hat B^{b}_{rs}(\vec x)]_{x_0=y_0}=
{2i\over g}\delta^{ab}\epsilon_{irs}\delta^{(3)}(\vec y-\vec x) \quad .
\label{tre.23}
\ea 
Owing to the cyclicity properties of the trace, (\ref{tre.19}) 
must be understood as symmetrized in $DB$ and $\Hol$.
Then, when (\ref{tre.21}) becomes operator valued,
the ordering 
procedure required by the quantization and enforced by (\ref{tre.18}), 
implies that we can substitute the product of fields with their
canonical commutator \cite{mmg}. 

Therefore using the Gauss theorem of (\ref{tre.18bis}) in (\ref{tre.19}), 
we express the divergence of the $B$ field as the flux through the
surface $S^2$ and obtain 
\be 
\hat M(C)|A> \simeq \tr \{ \unita+2ik\oint_C dy^i
\int_{\Sigma\sim S^2}d\sigma^{rs}_{(x)}\epsilon_{irs}
\delta^{(3)}(\vec y-\vec x)\delta^{ab}T_aT_b\} |A>\quad .
\label{tre.26}
\ee 
Notice now that
\ba 
&&\oint_C dy^i
\int_{\Sigma}d\sigma^{rs}_{(x)}\epsilon_{irs}
\delta^{(3)}(\vec y-\vec x) =
-\oint_C dy^i
\int_{\Sigma}{d\sigma^{rs}_{(x)}\over 4\pi}\epsilon_{irs}
{\pa\over\pa x^l} \biggl[ {\pa\over\pa x_l} 
\biggl( {1\over |\vec y-\vec x|} 
\biggr) \biggr]
\nonumber        \\
  &\simeq & {1\over 4\pi}\oint_C dy^i\oint_{C^{\prime}} 
dx^r\epsilon_{irl} 
{\pa\over\pa x_l}\biggl( {1\over |\vec y-\vec x|}\biggr) =
\Link(C,C^{\prime})\quad ,
\label{tre.27} 
\ea 
where $\Sigma=\Sigma^{\prime}\cup\Sigma^{\prime\prime}, \quad
\pa\Sigma^{\prime}=C^{\prime}$ and $C^{\prime}$ encloses the singularity of 
(\ref{tre.27}). Here we have used the formulae
\ba 
&& {\pa\over\pa x^l} {\pa\over\pa x_l} \biggl( {1\over |\vec y-\vec x|} 
\biggr) =
-4\pi  \delta^{(3)}(\vec y-\vec x)\quad ,
\nonumber \\ 
&& \int_{\Sigma}d\sigma^{rs}(x)\simeq
\delta\sigma^{rs}(\Sigma^{\prime}) =\oint_{C^{\prime}}dx^r x^s\quad ,
\nonumber    \\ 
\ea 
and the definition of the so called linking number between two closed loops, 
\ba 
\Link(C,C^{\prime}) &=& 
{1\over 4\pi}\oint_C dy^i\oint_{C^{\prime}} dx^r\epsilon_{irl} 
{\pa\over\pa x_l} ( {1\over |\vec y-\vec x|} ) = 
\label{tre.27bis} \\ 
&=&{1\over 4\pi} \oint_C dy^i\oint_{C^{\prime}} dx^r\epsilon_{irl} 
{(y-x)^l\over |\vec y-\vec x|^3}\quad ,
\nonumber 
\ea 
where $|\vec y-\vec x|$ is the usual Euclidean distance between two points. 
Notice that in our construction the closed contour $C^{\prime}$ is a framing 
contour of $C$, and hence the above linking number is the so-called  
self-linking number of $C$ \cite{rom}, 
\ba 
&& \sLink(C)\equiv \Link(C,C^{\prime})_{\epsilon\rightarrow 0}\quad ,
\nonumber      \\ 
&& C\equiv \{ x^i(t)\}\quad ,
\label{tre.28} \\ 
&& C^{\prime}\equiv \{ y^i(t)=x^i(t)+ 
\epsilon n^i(t)\ ;\ \epsilon >0,\ |n^i|=1\}  \quad ,
\nonumber 
\ea 
where $n^i$ is a vector  field orthogonal to $C$. 
The integral (\ref{tre.28}) is well defined and finite, 
takes integer values and 
equals the number of windings of $C^{\prime}$ around $C$. 

Putting the above formulae in 
(\ref{tre.26}), we find 
\ba 
\hat M(C)|A> &=& |A+D\varphi_C>=
\label{tre.30} \\ 
&\simeq& \tr \{\unita ( 1+2ikc_2(t)\sLink(C) ) \} |A>\quad ,
\nonumber 
\ea 
where 
\be 
\delta^{ab} T_aT_b=c_2(t)\unita\quad ,
\ee 
and   
\be 
c_2(t)={N^2-1\over 2N}\quad 
\label{tre.31} 
\ee 
for the fundamental representation of $SU(N)$. 
Eq. (\ref{tre.30}) implies that $\hat M(C)$ generates an infinitesimal 
multivalued gauge transformation $\varphi_C(\vec x )$. 
Whenever $\vec x\equiv\{ x^i\}$ traces a framing contour 
$C^{\prime}$ of the closed curve $C$ that winds 
$n\equiv \sLink(C)$ times around $C$, $\hat M(C)$ 
creates a magnetic flux \cite{thooft1}
\ba 
\Phi_C\equiv {k(N^2-1)\over gN}\sLink (C)\quad .
\label{tre.32} 
\ea 

At this order of approximation the finite multivalued
gauge transformation $\Omega_C[\vec x]$ generated  by the action of 
$\hat M(C)$ over some
state functional is given by
\be 
\hat M(C)|A(\vec x)>=|\Omega_C^{-1}[\vec x](A(\vec x)+id_{\vec  x})
\Omega_C[\vec x]>\simeq \tr \{ e^{ig\Phi_C}\unita\} |A(\vec x)> \quad .
\label{tre.33} 
\ee 

Because of the multi-valued nature of $\Omega_C[\vec x]$, 
one has that 
\be 
\Omega_C[\vec x_f]={\cal Z}_{\Phi_C}\Omega_C[\vec x_i]\quad ,
\ee
where 
$x^l_f\equiv x^l(t=1)=x^l(t=0)\equiv x^l_i$ is the base point of the loop 
$C$ parametrized by $C\equiv\{x^l(t)\ :\ 0\leq t\leq 1\} $ and 
${\cal Z}_{\Phi_C}$  is 
\be 
{\cal Z}_{\Phi_C}=e^{(ig\Phi_C)}\unita= 
e^{ik{(N^2-1)\over N}\sLink(C)}\unita\quad .
\label{tre.34}
\ee
Since $A^{\Omega_C}\equiv \Omega_C^{-1}(A+d)\Omega_C$ should always be 
single valued, ${\cal Z}_{\Phi_C}$ must be in the center of $SU(N)$.
To recover the standard form of the center,
we normalize  the free expansion parameter as $k=2\pi/(N^2-1)$
and require a special framing for $C$, namely that 
$n=\sLink(C)\in[0,\dots,N-1]$ .
With these normalizations the form of the color magnetic flux is given by
\be 
\Phi_C={2\pi n\over Ng}\quad .
\label{tre.38} 
\ee 

Given the properties of $\hat M(C)$, its commutation relations with the 
Wilson line
operator $\hat W (C_1)$ are easily deduced \cite{thooft1,ccgm,huang}: 
\be 
\hat M(C_1)\hat W(C_2)=e^{ig\Link(C_1,C_2)\Phi_{C_1}}
\hat W(C_1)\hat M(C_2) \quad .
\label{tre.41} 
\ee 

The previous equation is the well known `t Hooft algebra. The non 
trivial commutation relation between the operators $\hat M$ and $\hat W$ 
set their duality correspondence in the sense of the order and 
disorder operators in statistical mechanics.

\section{Computation of $<M(\Sigma , C)>$}
\setcounter{equation}{0}

The purpose of this section is to compute the average of the BF-observable 
$M(\Sigma ,C)$ defined in Sec. 3. 
Namely we define the dual loop functional by the normalized 
connected expectation value 
\be 
<M(C)>_{conn}\equiv {<M(C)> \over <1>}\quad ,
\label{qua.1}
\ee 
where 
\ba 
M(C)\!\! & \!\! \equiv &\!\!\!\! M(\Sigma ,C)\equiv \tr [ \exp \{ 
2\pi iqg\int_{\Sigma} 
d^2y\Hol^y_{\bar x}(\gamma )B(y) \Hol^{\bar x}_y (\gamma^{\prime} )\} ] \quad ,
\label{qua.2} 
\\ 
<1> \!\! & \!\! \equiv &\!\!\!\! \int {\cal D} A {\cal D}B 
\exp\{ -\int \tr [iB\wedge F +{g^2\over 4} B\wedge *B]\}\quad .
\label{qua.3} 
\ea 
With respect to the notation employed in the previous section, we have defined 
the expansion parameter $k$ in units of the bare 
color charge $g$ with a suitable normalization: $k=2\pi qg$.
\footnote{
For the sake of generality we do not ask any flux quantization and, as a 
consequence,
any fixed $k$ value. After 
gauge  fixing and a saddle point evaluation of the functional integral this 
quantization
will naturally emerge.}

Due to its $SU(N)$ gauge invariance, (\ref{qua.1})
must be gauge fixed. Following
Ref.\cite{thooft2} we choose the abelian projection gauge.  Before using this 
gauge 
we would
like to make  some comments. Magnetic monopoles are believed to play a major 
role in
confinement. They appear as classical solutions of the Yang-Mills theory 
in the  Georgi-Glashow model coupled with scalar fields transforming
in  the adjoint representation. Their classical mass is calculable, but 
turns out to be
too big to allow for a magnetic Higgs mechanism. The Abelian 
projection gauge
is supposed to generate monopoles of low or zero mass without introducing 
the
scalar Higgs field. As monopoles transform as $U(1)$ gauge fields, the idea 
is to
isolate the $U(1)^{N-1}$ degrees of freedom from those transforming in the 
coset
$SU(N)/U(1)^{N-1}$ with an appropriate partial gauge fixing leaving the 
$U(1)^{N-1}$
maximal abelian torus unbroken. The residual $U(1)^{N-1}$ gauge 
invariance
does not protect charged degrees of freedom from getting a mass
\footnote{
This mass does not have to be physical and it can also be an infrared 
regulator.
The decoupling of charged degrees of freedom is insensitive to the 
nature of 
the mass.}.
Since confinement is a large scale phenomenon, in the crudest of 
approximations,
these degrees of freedom can be discarded. Therefore the effective theory 
should contain only ``photons" and monopoles.

To implement this gauge we need a microscopic field or a 
composite of it, $X$, to
transform in the adjoint representation of the gauge group 
$SU(N)$ so that 
a gauge
transformation can diagonalize it,
\be
\tilde X=g^{-1}Xg={\rm diag}(\lambda_1,\ldots,\lambda_N)\quad .
\label{qua.4}
\ee
The eigenvalues of such matrix can be naturally ordered
$\lambda_1\geq\lambda_2\geq\cdots\geq\lambda_N$. When two eigenvalues
coincide we have a line of singularities representing the monopole
world-line \cite{thooft2}. By analogy with the interpolating gauges used in 
massive
Yang-Mills \cite{thooft3}, having chosen (for example) $X=F_{12}$ and
$Y_{ij}={(F_{12})_{ij}/(\lambda_i-\lambda_j)},
\quad i\neq j=1,\ldots,N$,
the proposed gauge condition is 
\be
Y^{ch}+\xi D^0*A^{ch}=0\quad ,
\label{qua.5}
\ee
where the superscripts $ch, 0$ stand for the off-diagonal and diagonal 
part of the matrix
$A=A^0+A^{ch}$ and $D^0$ is the covariant derivative with respect to the 
diagonal part of the gauge field. Interpolating gauges are such that for 
short
distances the relevant gauge is $ D^0*A^{ch}=0$ and the theory is 
renormalizable,
while for large distances the gauge is $Y^{ch}=0$. For this 
reason the field
$Y$ needs to be adimensional and $\xi$ has the dimension of 
the inverse of a 
mass squared.  In standard YM theory, $X$ can only be a composite of 
the $F_{\mu\nu}$ 
(and its
covariant derivatives) in order to transform according to the adjoint 
representation.
But as the dependence of $F_{\mu\nu}$  on the momentum $p_\mu$ 
is the same 
as the
second term in the r.h.s. of (\ref{qua.5}) the dominance argument
is only possible if $X$ is made adimensional. This makes this 
gauge difficult to 
apply to standard quantum Yang-Mills theory because $Y$ 
is non-polynomial
outside the tube around the monopole string and divergent inside it.

Quite remarkably, these problems are not present in the $BF-YM$ theory
due to the presence of the microscopic $B$ field. The 
interpolating gauge
can now be easily implemented choosing, for instance, $X=B_{12}$:
\be 
B^{ch}_{12}+D^0*A^{ch}=0\quad .
\label{qua.6} 
\ee 
Equivalently in our formalism we can diagonalize the two-form $B$ 
on the surface $\Sigma$
\ba 
&\tilde B&=V^{-1}BV=diag(\beta_1 ,...,\beta_N )\ ,
\quad \beta_1\geq\beta_2 \geq\ldots\geq\beta_N \quad ,
\label{qua.7} \\ 
&\tilde A&=V^{-1}(A+d)V\ ,
\quad \tilde F=V^{-1}FV  \quad ,
\nonumber 
\ea 
and then use the background gauge condition $D^0*A^{ch}=0$ in the 
renormalization program.

According to the previous observations, in the computation of  the 
expectation value of 
the observable $M(C,\Sigma)$, employing  the abelian
projection gauge, in the large scale region we may neglect 
the massive off-diagonal degrees of freedom $B^{ch}, A^{ch}$. This approximation
is often called Abelian dominance (\cite{thooft2,eza}).

We would like now to interpret this approximation from a different
point of view which will turn out to be very useful in the following.
The existence of monopoles in the Abelian projection gauge is due
to the compactness of the $U(1)^{N-1}$ group and it is related to
the existence of non trivial topological objects for the entire $SU(N)$
theory. So, in this contest,
 Abelian dominance can also be seen as the prescription 
of having reducible gauge connection which, as we will find later,
will also be singular. In the simple $SU(2)$ case this leads
to a connection
\be
A={1\over 2}\pmatrix{\alpha&0\cr 0&-\alpha}\quad .
\label{qua.7bis}
\ee
The reducibility of the $SU(2)$ gauge connection, implies that the
gauge bundle is split and thus
requires the existence of a positive definite first Pontrjagin class and 
intersection number (which we will define later on). 
This fact, in its turn, implies the absence of anti-self-dual harmonic
(closed) two forms \cite{freed}. Later on in this chapter, we will comment
on the possible relation between this split bundle and the
moduli space of instantonic classical solutions.

In the general $SU(N)$ case
we now rewrite the functional integral in terms of the variables
\be 
\alpha_i=[A^0]_{ii}\ ,\quad i=1,...,N,\quad \sum_i\alpha_i=0 \quad ,
\label{qua.8} 
\ee 
and
\be 
f_i\equiv [F^0]_{ii} =d[A^0]_{ii}\qquad i,j=1,...,N \quad .
\label{qua.9} 
\ee 

It is then 
convenient to rescale the previous geometrical fields 
$A^0$, $B^0$ to the physical ones 
$A^0\rightarrow gA^0$,  
$B^0\rightarrow {1\over g}B^0$. 
Furthermore, we replace the surface integral in Eq. (\ref{qua.2}) with the 
integration over the so-called Poincar\`e dual form, $\omega_{\Sigma}$, 
of (the homology class of ) the surface $\Sigma$ \cite{ccgm}.
By definition, $\omega_{\Sigma}$ is closed.
Moreover, choosing an orientation of the four manifold
and remembering the absence of anti-self-dual harmonic two forms 
(which is imposed by 
requiring the existence of a gauge connection of the type of (\ref{qua.7bis})) 
$\omega_{\Sigma}$ is chosen to be self-dual {\it i.e.} 
$*\omega_{\Sigma}=\omega_{\Sigma}$ and  
with the property that (up to gauge one-forms) 
for a generic two-form $t$ 
\be 
\int_{\Sigma} t\simeq \int_{M^4}\omega_{\Sigma}\wedge t \quad .
\label{qua.10}
\ee  
In  a local system of coordinates  $(x,y,u,v)$ on $\Sigma$, so that $\Sigma$ 
is given by the equations $x=y=0$, the dual form $\omega_{\Sigma}$ can be taken as 
$\omega_{\Sigma}\simeq \delta^{(2)}(x,y)dx\wedge dy$ 
and normalized as 
\be 
\int_{N(\Sigma )}\omega_{\Sigma}\simeq \int \delta^{(2)} (x,y) dx\wedge dy =1
\quad ,  
\label{qua.11}
\ee 
where $N(\Sigma )$ is the transversal tubular neighbourhood on the surface 
$\Sigma$. 

Let us now compute the average of the `t Hooft loop operator $M(C)$ in the 
above abelian projection scheme. 

The functional integrations over the 
$\alpha_i$'s should be constrained by $\sum_i\alpha_i =0$. 
In order to perform the calculations 
without imposing  
the previous constraint, 
we extend the 
$SU(N)$  gauge group of the BF-YM theory to $U(N)$ \cite{thooft4}. 
More precisely we identify the fields
$\alpha_i$ with  the Cartan generators of $U(N)$.
Then we can recover the original $SU(N)$ BF-YM theory by 
gauging the spurious $U(1)$ symmetry, generated by 
$\alpha_N$, associated to the standard embedding 
$SU(N)\rightarrow U(N)/SU(N)=U(1)$. Finally the abelian 
gauge fields $\alpha_1,\ ...,\ \alpha_{N-1}$ associated to the 
Cartan subalgebra of $SU(N)$ should be gauged to remove the 
remaining $U(1)$ gauge degrees of freedom. 

Using the Abelian dominance approximation, 
the `t Hooft loop operator $M(C)$ becomes 
\ba 
&& M(C)={\rm Tr}\{ O_{ij}(C)\} \quad ,\label{qua.12} \\ 
&&O_{ij}(C)=\delta_{ij}\exp \{i2\pi q\int_{M^4}\omega_{\Sigma} \wedge
\beta_j [\cos (g\oint_C \alpha_j ) +i\sin (g\oint_C \alpha_j )]\} \quad . 
\nonumber 
\ea   

As we have already observed, the gauge transformation in (\ref{qua.7}) is 
singular \cite{thooft2}, and the singularities occur if two consecutive 
eigenvalues of $B$ coincide. If $\beta_i =\beta_{i+1}$, we shall  
label such a point by $x^{(i)}$. Moreover $\beta_i$ is a two-form and hence 
its support is a two-cycle in the four manifold. Thus, in order to avoid the 
previous singularities $x^{(i)}$, we must remove a ball 
$S^2_{\epsilon}(x^{(i)})$ of radius $\epsilon\rightarrow 0^+$ at each point 
$x^{(i)}$ of the base manifold. 
As a consequence  of that, we get a non trivial 
Stokes theorem ($g\alpha_i$ is the geometrical $U(1)$
connection ):
\be 
g\oint_C\alpha_i =g\int_{S\ :\ \pa S=C}d\alpha_i
=g\int_{S^2_{\epsilon}(x^{(i)})}d\alpha_i\equiv 2\pi gq_i
\label{qua.13}
\ee 
$(S=S^{\prime}\cup S^2_{\epsilon}(x^{(i)}))$, since the two-cycle $S$ is not 
homotopically trivial (it contains $S^2_{\epsilon}(x^{(i)})$). 
In (\ref{qua.13}) the magnetic charge is given by the first Chern class, 
$c_1(L)$, of the line
bundle  $L$ with fiber $S^2_{\epsilon}(x^{(i)})$. Since we have in principle 
$(N-1)$ connections $U(1)$ whose curvatures are supported
near the points $x^{(i)}$ of 
$\Sigma$ ($x^{(1)}$ is identified with $x^{(N)}$) 
and the total first Chern class is an 
integer ($c_1(L^{\otimes (N-1)})=(N-1)c_1(L)\in\zet$), we obtain 
that $gq_i$ satisfies the Dirac quantization condition rule 
\be 
gq_i={n_i\over N-1}\quad ,\quad n_i=\pm 1,\ \pm 2,\ ...\quad ,\quad 
\forall i\in [1,N] \quad .
\label{qua.14}
\ee 
In other words the appropriate homotopy 
group is 
$\Pi_2 [SU(N)/U(1)^{N-1}]=\zet^{N-1}$, as expected.

In the following we shall show that the effective field 
equations associated to the strong coupling limit of 
$\Gamma (C)\equiv -\ln <\tilde M(C)>_{conn}$ admit as 
critical points magnetic monopole strings - which we shall construct 
explicity - with winding numbers $n_i\in\zet$ at the singular points 
$x^{(i)}$.  
The key point will be the identification of 
the arbitrary expansion parameter $q$ in (\ref{qua.2}) with 
the magnetic charges $q_i$, i.e.  we will set 
\be 
q_i\propto q\propto {1\over g}\quad .
\label{qua.15}
\ee  
We start now the computation of the magnetic order 
parameter (\ref{qua.1})  in the $q\rightarrow 0$, which will be identified 
by (\ref{qua.15}) as the strong coupling limit $g\rightarrow\infty$. 
In this 
limit the operator (\ref{qua.12}) can be approximated by 
\footnote{We note that the appearence of the effective expansion 
parameter $q/N$ in
(\ref{qua.16}) suggests that the same conclusions of this section 
can also be obtained in the 
large $N$-limit (holding $q$ fixed) in which the $SU(N)$ and the $U(N)$ gauge 
theory
are the same \cite{thooft4}. Thus, in the $q\to 0$ limit, 
we can consistently drop the constraint $\sum_i\alpha_i =0$.
}
\be 
M(C)=\sum_i O_{ii}(C) \simeq 
N\exp \bigg\{{i2\pi q\over N}\int\omega_{\Sigma}\wedge
\sum_i\beta_i\cos (g\oint_C\alpha_i )\bigg\}\quad .
\label{qua.16}
\ee 
Eq. (\ref{qua.16}) is obtained by taking into account that, according to our 
previous observations, there are two types of monopole configurations 
$\alpha_i$ for each $i$, depending on the sign of the magnetic charges $q_i$. 
Indeed we have that 
$\{ q_i\}\equiv\{ q_i^s\}\equiv \{ q_i^+>0,q_i^-<0\}$. Hence it is resonable 
to define the sum over the field configurations as $\sum_i\equiv\sum_{i,s}$, 
where $i\in [1,N]$ and $s=\pm$. 
Notice that 
$<M(C)>_{conn}$ in the strong coupling limit is the partition function of 
$N$ compact QED's or 4D Villain models (e.g. see \cite{poly}). 

With all these approximations taken into account the form of the magnetic
order parameter in the strong coupling region becomes
\ba
&&<M(C)>_{conn} \!\!\!\!\!\!\!\! 
\simeq {N\over <1>}
\int {\cal D}\alpha_i {\cal D}\beta_i 
\exp \{-{1\over 2}\sum_i\int ({1\over 4}\beta_i\wedge *\beta_i 
+i\beta_i\wedge [d\alpha_i 
\nonumber  \\
&&+{4\pi q\over N}
\omega_{\Sigma}\cos (g\oint_C\alpha_i )])\} = {N\over <1>}
\int {\cal D}\alpha_i  
\exp \{-{1\over 4}\sum_i\int  [d\alpha_i +{4\pi q\over N}
\omega_{\Sigma}\cos (g\oint_C\alpha_i )]^2)\}\quad , 
\label{qua.17}
\ea
where the square of a form $t$ means $t\wedge * t$.
We now split the gauge fields as $\alpha_i =\bar\alpha_i +\hbar Q_i$, 
where  the quantum fluctuations $Q_i$ must be gauged (e.g. by a covariant 
gauge condition) 
and the $\bar\alpha_i$ are singular classical configurations.  
Postponing for a while the discussion of quantum fluctuations,
we concentrate on the semi-classical contribution to the path
integral which is
\be 
<M(C)>_{conn}\simeq N\exp \{-{8\pi^2 q^2\over N^2}\sum_i
\int \omega_{\Sigma}\wedge\omega_{\Sigma^{\prime}}
\cos (g\oint_C\bar\alpha_i )
\cos (g\oint_{C^{\prime}}\bar\alpha_i^{\prime} )\}\quad .
\label{qua.18b} 
\ee 
In the derivation of the above equation, we have exploited the 
following facts:
\begin{itemize}
\item
partial integration is not allowed on the $\bar\alpha_i$ due 
to their singular behaviour \cite{kron}, 
\item
$\bar\alpha_i$ obeys the
(monopole) equation
\be 
*d\bar\alpha_i={4\pi q\over N}\omega_{\Sigma}
\cos (g\oint_C\bar\alpha_i)\quad ,
\label{qua.19}  
\ee
\item 
the absence of electric currents in the model,
\item 
the self-duality and closedness of $\omega_\Sigma$.
The above properties 
imply the absence of terms linear in $Q_i$. 
\end{itemize}

Equations of the type (\ref{qua.19}),  appeared already
in the study of the duality properties of 
gauge theories and 4D manifold invariants \cite{fre,witmono}. 

The r.h.s. of (\ref{qua.19}) can be written in terms of 
spinor fields 
for a direct comparison with Ref.\cite{witmono}. 
If $M^4$ is a spin manifold any  self  dual two-form $B^+$, 
satisfying the conditions (in spinorial notation) 
$B^+_{(AB}\wedge B^+_{CD)}=0$ and $B^{+AB}\wedge B^+_{AB}\neq 0$, 
can be  represented  
in the form $(B^+)^{AB}={1\over 2}(M_A \bar M_B +M_B\bar M_A )$, 
where $M$ is 
a two component spinor field.  
Furthermore $B^+_{\mu\nu}\eta^{a}_{\mu\nu} {1\over \sqrt 2}
\sigma^a_{AB}=B^+_{AB}$, 
where $\eta$ are the `t Hooft symbols and $\sigma$ the Pauli matrices. 
In terms of the space-time indices 
$B^+_{\mu\nu}=\bar M\Gamma_{\mu\nu}M$, with 
$\Gamma_{\mu\nu}={1\over 2}[\gamma_{\mu},\gamma_{\nu}]$. 
Using the equation of motion for the $B$ field, $M$ is seen to obey
the Dirac equation relative to the case of manifolds which are
K\"ahler and spin as discussed in section four of Ref.\cite{witmono}.
This connection with Ref.\cite{witmono}, in which the infrared limit 
of the supersymmetric $N=2$ theory is described in terms of abelian 
gauge fields and monopoles, should not be surprising. 
In fact, if we take the attitude (that will be developed in the
next section) that the term $g^2/4 B\wedge *B$ is a perturbation
of the ``pure BF" theory (\ref{uno.4}), this perturbation theory
will be performed around the vacuum of the theory which 
coincides with that of (\ref{uno.4}).
This requires to gauge fix the local symmetries (\ref{tre.6}), 
(\ref{uno.5}) of the action (\ref{uno.4})
by introducing appropriate ghosts \cite{blau}.

We would now like to explain in which respect the pure BF
theory (with action (\ref{uno.4})) deserves to be called ``topological".
In first place the action (\ref{uno.4}) is metric independent.
Moreover, after some work, a BRST charge , $s$ (nihilpotent
on-shell) and a vector charge, 
$\delta_\mu$, operator can be defined for the quantized theory \cite{blau}.
The commutator $\{s,\delta_\mu\}$ closes on translations 
on-shell and this 
shows that the stress energy tensor of the theory can be written as a BRST 
commutator. Introducing external sources coupled to the BRST
variations of the microscopic fields a new linearized BRST 
charge, $\Omega$ (nihilpotent off-shell), can be defined. 
Thus the BF theory is proved to be ``cohomological" \cite{gms} with
respect to the $\Omega$ operator but it is not clearly equivalent
to the topological theory defined in \cite{wittopo} since the BRST 
charges of the two theories are different. 

Independently from these considerations, the symmetries (\ref{tre.6})
must be anyway gauge fixed, which can be done using a 
covariant gauge for the field $B$. 

We now look for an explicit solution of (\ref{qua.19}) under the 
constraint $\cos (g\oint_C\bar \alpha_i )=1$ which
is consistent with the Dirac quantization rule (\ref{qua.14}) if we set 
$n_i = m(N-1),\quad m\in\zet,\quad q_i={2\over N}q$. 
We then get 
\be 
qg={mN\over 2}\quad .
\label{qua.20}
\ee 

Given an arbitrary and topologically trivial surface $\Sigma$ the dual
Poincar\'e form can be locally written as 
\ba
\omega^\Sigma_{\mu\nu}(x)&=&\int_\Sigma d\sigma_{\mu\nu}(y)
\delta^{(4)}(x-y)
\quad ,\label{qua.21}\\
d\sigma_{\mu\nu}(y)&=&\varepsilon^{ab}{\partial 
y^\mu(\xi)\over\partial\xi^a}
{\partial y^\nu(\xi)\over\partial\xi^b}d^2\xi=\varepsilon^{ab}
t^\mu_at^\nu_bd^2\xi\quad ,
\nonumber
\ea
where $\xi_a$ are the coordinates on the surface, $t_a^\mu$ 
are the normal vectors
of the bidimensional surface $\Sigma$ and $a=1,2$. Putting 
(\ref{qua.21}) 
in (\ref{qua.19}) we find
the classical background solution \cite{lee,msz}
\be
\bar\alpha^i_{\mu}(y)=4q_i \int_{\Sigma}d^2\xi 
P^+_{\mu\nu\alpha\beta}
\varepsilon^{ab}t^\nu_at^\alpha_b\partial^\beta 
{1\over |y-x^{(i)}|^2}\quad ,
\label{qua.22}
\ee 
where $P^+=(1+*)/2$ and, as before, $x^{(i)}$ is the location of 
the singularity at the intersection
of the surface $\Sigma$ with the $i$-th monopole world-line and $\Sigma $ 
locally appears as the ``Dirac sheet'' \cite{thooft1}. The curve $C$ of 
(\ref{qua.19})
clearly winds around $x^{(i)}$.
Eq. (\ref{qua.22}) is a magnetic vortex-line or Abrikosov-Nielsen-Olesen 
(ANO) string for each $i\in [1,N]$ \cite{no}, obtained here without 
the use of the Higgs  
fields, but introducing the disorder effect 
of the $M(C)$ operator.
\footnote{
In Ref.\cite{polykarpo,no} the same solution is found in the abelian
Higgs model with $\bar \alpha^i_{\mu}=q_i\partial_{\mu}\chi$, 
and the Higgs field is parametrized  as $\varphi =|\varphi |
\exp (i\chi )$. 

In Ref.\cite{msz} these kind of configurations are shown to describe 
massless monopoles in 4D compact QED, giving rise to a planar model like  
thermodynamics.}
We stress that these configurations do not correspond to  
`t Hooft-Polyakov like monopoles but are rather singular Dirac 
monopoles \cite{freund}. 
The bidimensional reduction of the solution (\ref{qua.22}) (obtained
fixing the surface $\Sigma$) coincides with eq.(31) of Ref.\cite{nst}
(near the singularity) which is a solution of the dimensionally
reduced monopole equations of Ref.\cite{witmono}.

Finally, the classical contribution is
\be 
S_0= { 2\pi^2 m^2\over g^2}\int\sum_i
\omega_{\Sigma}\wedge\omega_{\Sigma^{\prime}}
\cos (g\oint_C\bar\alpha_i) 
\cos (g\oint_{C^{\prime}}\bar\alpha^{\prime}_i) \equiv 
{ 2\pi^2 m^2N\over g^2}Q(\Sigma,\Sigma^\prime)\quad .
\label{qua.26}
\ee 
For (\ref{qua.26}) to make sense, the 
two cosines must be different from zero which implies 
that the paths $C, C^\prime$ must encircle monopole flux tubes.
There are now two options:
\begin{itemize}
\item
$g\oint_C\bar\alpha_i\neq g\oint_{C^\prime}
\bar\alpha^\prime_i$
\item
$g\oint_C\bar\alpha_i= g\oint_{C^\prime}
\bar\alpha^\prime_i$\quad .
\end{itemize}

In the first case, $Q$ is the intersection number \cite{donald}
\be 
Q(\Sigma,\Sigma^\prime)=\int_{M^4} \omega_{\Sigma [C]}\wedge 
\omega_{\Sigma^{\prime} [C^{\prime}]} \quad ,
\label{qua.27}
\ee 
which is an integer (it is a four dimensional topological invariant) 
characterizing the self-intersection of $\Sigma$ with itself.
This number
is given once the surface $\Sigma$ and the ambient space $M^4$ are assigned.

In the second case the 
closed curve $C^{\prime}\equiv \{ y^{\mu} (t)\} $ is a framing contour 
of the closed curve $C\equiv \{ x^{\mu} (t)\} $, i.e. iff it happens that 
\be
y^{\mu}(t)=x^{\mu}(t) +\epsilon n^{\mu}(t)\ ,\ \epsilon\to 0\ ,\ |n^{\mu}|=1
\quad ,
\label{qua.28}
\ee 
where $n^{\mu}(t)$ is a vector field orthogonal to $C$. 
In this case $Q(\Sigma,\Sigma^{\prime})$ becomes the  
self-linking number of $C$ given by (\ref{tre.28}).

Then we get that 
\be 
Q(\Sigma,\Sigma^\prime)= \sLink(C)\quad .
\label{qua.30}
\ee 

From a physical point of view, we may define $\sLink (C)$ 
in (\ref{qua.30}) as 
\be 
\sLink(C)\equiv {L(C)\over \rho}\quad ,
\label{qua.31}
\ee 
where $\rho$ plays the role of unit of 
lenght and $L(C)$ is the perimeter of the 
loop $C$.
We may understand (\ref{qua.31}) by a n-vertex polygon discretization 
$C\to C_{PL}$ of the loop $C$. Indeed by the very definition of the 
self-linking number one gets that 
$\sLink (C_{PL})=n
=L(C_{PL})/l$, 
where $l$ is the lattice spacing. (\ref{qua.31}) is obtained in the 
continuum limit $n\to\infty, l\to 0$, 
since in this limit $L(C_{PL})/l\sim L(C)/\rho $. 
Thus, in this picture, the perimeter law comes out from the arbitrary 
framing dependence 
of the path $C$ in the color magnetic operator $M(C,\Sigma)$. In turn, 
the framing dependence can be seen as a point
splitting regularization due to the non-local nature of the operator.
In our construction $\sLink(C)$ in 4D  cannot be slinked - in general the 
linking between two arbitrary curves in 4D has no geometrical meaning - 
since $C$ and $C^{\prime}$ belong to the same surface.

Let us now discuss quantum fluctuations. If the effective theory for
large length scales is a $U(1)$ type theory, for short length scales 
the charged
degrees of freedom cannot be discarded anymore. Let $\Lambda$ be the 
length scale 
separating these two regimes and let us divide the gauge field 
according to a background field prescription: 
$A^a=\bar A^a+Q^a$ where $a$ is a gauge index. 
Moreover the gauge field can be written
as the sum of two parts giving the contribution of the large length scale 
sector
($>\Lambda$) and the short length scale sector ($<\Lambda$); the functional 
integration
over the gauge field factorizes in a way compatible with this separation.
Moreover let, for the sake of simplicity, the gauge group be $SU(2)$. 
For length scales bigger than $\Lambda$ we take $A^3=\bar\alpha+Q^3$ 
which is the 
usual $U(1)$ prescription. For length scales smaller than $\Lambda$ we take 
$A^a=\delta^{a3}\bar\alpha+Q^a$, {\it i.e.} we continue the
classical solution into the small scales region where the quantum 
fluctuations coming from the charged degrees of freedom cannot be
discarded. The expectation is that the small scales behaviour is
insensitive to the classical solution according to the background field
method. Performing the functional integration over the quantum
fluctuations leads to a double contribution, in complete analogy
with the saddle point evaluation around an instanton background 
\cite{mrso}.
\begin{itemize}
\item
The first contribution is given by a ratio of determinants given by
\be
\left[ {Det^\prime (-L_0)
\over Det^\prime(-L)}
\right] ^{1\over 2}\left[ {Det(-\bar D^2)\over Det(-\partial^2)}
\right]\quad ,
\label{qua.31bis}
\ee
where $\bar D=d -ig\bar\alpha$, $\bar D^2\equiv\bar 
D^{\mu}\bar D_{\mu}$,
\be 
L=\bar D^2\delta_{\mu\nu} -(1-{1\over\xi})\bar D_\mu\bar D_\nu
+2ig\bar F^0_{\mu\nu}(x) \quad ,  
\label{qua.31ter}
\ee
and $L_0$ is given from $L$ evaluated around the 
trivial background.
$\xi$ is a gauge parameter usually chosen to be one. 
In (\ref{qua.31bis})
the determinants are primed to remind the reader 
that the contribution
of zero modes is omitted and that the determinants are 
regularized.
\item
The second contribution is given by the Pauli-Villars 
regularization 
of the determinants and it amounts to a scale $\mu$ 
(which is the 
Pauli-Villars mass) raised to a certain power which 
is given by 
the dimension of the moduli space of the 
classical solution.
\end{itemize}

Let us now proceed with the evaluation of these two 
contributions.
Using the self-duality property of our classical solution, 
the ratio
of determinants (\ref{qua.31bis}) can be written as
$R=\left[ Det(-\partial^2)/Det(-\bar D^2)\right]$ \cite{mrso}. This 
ratio has been
evaluated in Ref.\cite{cgot} using the heat kernel method 
in the case
of an $SU(N)$ gauge group but it is easy to generalize 
this result to our case too. We now give a brief
outline of this  computation to justify the previous statement.
 
To be mathematically well-defined, the split connection (\ref{qua.7bis}) 
must be considered either
to be regular in $\Re^4$ or to be singular in $S^4$.
In our case  we found explicitely a singular $U(1)$ connection
and we can go from one point of view to the other by removing the
singularity locus spanned by the $x^{(i)}$ in (\ref{qua.22}). 
Indeed, by removing a disk from $S^4$ we in fact obtain $\Re^4$.

The computation we want now to perform is intended to probe the region
around the singularity by evaluating the operator $R$ obtained by
expanding around the classical solution given in (\ref{qua.19}).
This computation is carried on a  space-time bounded by the length 
scale $\Lambda$
which can be taken to be the compact manifold $S^4$ after a conformal
transformation, 
to allow a proper definition of the eigenvalue problem and to avoid 
infrared problems. 

The starting point of the computation
is the introduction of the 
Riemann zeta function $\zeta(s)$ built out of the eigenvalues of 
the operator $-\bar D^2$.
The logarithm of the ratio of the determinants of the self-adjoint 
operators $R$ is given by the variation, 
under conformal transformations, of the derivative of the Riemann zeta 
function $\zeta^\prime(0)$ continued to zero value of its argument.
In turn this function is given by the residue of 
\be
2\int_0^\infty t^{s-1}\tr\lbrack G(x,x,t)\delta\omega\rbrack dt
\label{qua.32}
\ee
at $s=0$ \cite{cgot}. In (\ref{qua.32}), $\delta\omega$ is an 
infinitesimal conformal transformation and 
\be
G(x,y,t)={1\over 16\pi^2 t^2}e^{-{(x-y)^2\over 4t}}
\sum_{n=0}^\infty a_n(x,y)t^n
\label{qua.32bis}
\ee
satisfies the heat equation.

Making use of (\ref{qua.32bis}), (\ref{qua.32}) can be cast in
the form
\be
R=\zeta^\prime(0)={1\over 8\pi^2}\int\tr a_2(x,x)\delta\omega d^4x\quad .
\label{qua.32ter}
\ee

Plugging the expansion (\ref{qua.32bis}) in the heat-kernel differential 
equation we obtain a set of recursion relations from which the values
of the coefficients $a_n$ are easily extracted. Since the only 
manipulation
needed in solving such recursion relations is the commutation of 
covariant derivatives, the final result is easily generalized to 
any gauge
group
\be
\ln{Det(-\partial^2)\over Det(-\bar D^2)}=
{\ln({\mu\Lambda})\over 96\pi^2}
\int \bar F^0_{\mu\nu}(x) \bar F^0_{\mu\nu}(x)d^4x=
{\ln({\mu\Lambda})\over 96\pi^2}
{1\over 4}\int (d\bar\alpha )_{\mu\nu}(x)  
(d\bar\alpha )_{\mu\nu}(x)d^4x\quad .
\label{qua.35}
\ee 
The factor 1/4 comes from the normalization of the gauge group generators
according to (\ref{qua.7bis}).

The contribution coming from the regularization of the zero modes 
is obtained
once the dimension of the moduli space is computed, according to 
Ref.\cite{witmono}, to be
\be
\dim {\cal M}=c_1(L)^2=c_1(L)\wedge c_1(L)={1\over 32\pi^2}\int 
(d\bar\alpha )_{\mu\nu}(x) (d\bar\alpha )_{\mu\nu}(x)d^4x\quad .
\label{qua.36}
\ee

Putting together the classical result (\ref{qua.26})
with the quantum fluctuations, 
we find that the bare coupling
$g$ can be substituted by its renormalized expression 
and that (\ref{qua.18b}) can be written as
\be
{8\pi^2\over g^2_R}{c_1(L)^2\over 8}={c_1(L)^2
\over 8}\biggl({8\pi^2\over g^2}
-(8-{2\over 3})\ln ({\mu\Lambda})\biggr)
\equiv {c_1(L)^2
\over 8}\biggl({8\pi^2\over g^2}-\beta_1\ln ({\mu\Lambda})
\biggr)\quad ,
\label{qua.37}
\ee
where $\beta_1={22\over 3}$ is the first coefficient of the $SU(2)$ 
beta function of the non-abelian Yang-Mills theory.

\section{The Average of the Wilson Loop }
\setcounter{equation}{0}

In this section we shall compute the average of the Wilson loop
and find an area law behaviour for its leading part. 
Furthermore, in our formalism, the area law gets a nice geometrical 
interpretation: it is the response of the true QCD vacuum to arbitrary 
deformations of the quark loop ${\cal C}$.

The starting point here is given by the Wilson loop operator written in 
terms of the non abelian Stokes theorem (see e.g. \cite{ara} ):
\be
W_R(\C) \equiv W_R(\Sigma ,C)=Tr_R\{ P_{\Sigma}
\exp [i\int_{\Sigma} \Hol_{\bar x}^x (\gamma )
F(x) \Hol^{\bar x}_x(\gamma^{\prime})]\}\quad ,
\label{cin.1} 
\ee
where $\C=\partial\Sigma$, $C=\{\gamma(x)\cup\gamma^{\prime}(x)\}$ 
was defined at
the beginning of section 2 and $P_{\Sigma}$ means surface path ordering.
$W_R$ is 
calculated with respect to some irreducible representation $R$ of $SU(N)$. 
The loop average in the fundamental 
representation $R\equiv t$ of $SU(N)$ is
\be 
<W_t(\C)>_{conn}\equiv {<W_t(\Sigma ,C )>\over <1>}
\equiv 
{\int {\cal D}B{\cal D}A \ W_t(\Sigma ,C )e^{-S_{BF-YM}}\over 
\int {\cal D}B{\cal D}A\ e^{-S_{BF-YM}}} \quad ,
\label{cin.2}
\ee 
where $S_{BF-YM}$ was defined in (\ref{uno.1}). 

Expanding in series the Wilson loop (\ref{cin.1}) we get 
\be 
<W_t(\Sigma,\C)>=\sum_n<{1\over n!}{\tr}_t 
P_\Sigma\int_{\Sigma_1}\ldots\int_{\Sigma_n}
(i\Hol(\gamma )F(x) \Hol^{-1}(\gamma^{\prime}))^n>\quad .
\label{cin.4}
\ee
We then use the identity 
\be  
e^{-\frac i 4 \int *B^a_{\mu\nu}F^a_{\mu\nu}} F^a_{\rho\sigma}(x)=
4i\frac {\delta} {\delta *B^a_{\rho\sigma}(x)}(e^{-\frac i 4 \int 
*B^a_{\mu\nu}F^a_{\mu\nu}})\quad .
\label{cin.6}
\ee  
Performing a partial integration with respect to the functional
derivative in (\ref{cin.6}) we can replace, in the path integral,
\be 
i\Hol(\gamma )e^{-\frac i 4 \int *B^a\cdot F^a}F(x) 
\Hol^{-1}(\gamma^{\prime})\rightarrow 
4\Hol(\gamma )e^{-\frac i 4 \int *B^a\cdot F^a}
({\delta\over \delta *B(x)})\Hol^{-1}(\gamma^{\prime})\quad .
\label{cin.7}
\ee 
The functional derivative acts only to its right on the exponential
of the mass term $-g^2/16\int B^a_{\mu\nu}B^a_{\mu\nu}$, since $\Hol$ does not
contain the $B$ field.
We now need the identity 
\be 
V(\frac {\delta} {\delta *B^a(x)_{\rho\sigma}})e^{-\frac {g^2} {16}
\int B^a_{\mu\nu}  
B^a_{\mu\nu}}=
V(-\frac {g^2} {8} *B^a_{\rho\sigma})e^{-\frac {g^2} {16}\int B^a_{\mu\nu}
B^a_{\mu\nu}}\quad ,
\label{cin.10}
\ee 
where $V$ is the functional defined by 
\be 
V(\frac {\delta} {\delta *B^a(x)})\equiv 
P_\Sigma\int_{\Sigma_1}\ldots\int_{\Sigma_n}
(4\Hol(\gamma ) 
(\frac {\delta} {\delta *B^a(x)}) 
\Hol^{-1}(\gamma^{\prime}))^n\quad .
\label{cin.11}
\ee 
Resumming the exponential series for the Wilson loop 
we finally get the ``duality" relation 
\be 
<W_t(\C)>_{conn}={<M^*_t(\Sigma ,C,\C=\partial\Sigma)>
\over <1>}\quad , 
\label{cin.9}
\ee 
where
\be 
M^*_t(\Sigma ,C,\C=\pa\Sigma)=
Tr_t[P_{\Sigma} \exp \{ -{g^2\over 2}\int_{\Sigma}
\Hol_{\bar x}^x (\gamma )
*B(x) \Hol^{\bar x}_x(\gamma^{\prime})\} ] \quad .
\label{cin.8}
\ee 
$M^*_t(\Sigma ,C)$ is the dual (in the sense that 
$B\to *B$) of the observable 
$M_t(\Sigma ,C)$ defined in (\ref{tre.2}) with $k$ set to $k=ig^2/2$.

To calculate (\ref{cin.9}) we expand perturbatively  in $g$ both 
the exponential and 
the holonomies which appear in the exponent of (\ref{cin.8}) \cite{cm}. 
The first relevant contraction 
encountered at lower level is given in terms of $<A*B>$, 
which can be computed starting from the off-diagonal 
propagator $<AB>$. The latter propagator is the same for 
the BF-YM theory \cite{mz} 
and the pure BF (\ref{uno.4}). Therefore we find 
\be 
<M^*_t(\C)>_{conn}  =e^{-{g^2\over 2} c_2(t)\oint_{C}\int_{\Sigma}<A*B>}
\Delta (\Sigma)\quad ,
\label{cin.13}   
\ee 
where $c_2(t)$ is given in (\ref{tre.31}) and $\Delta (\Sigma)$ depends 
on higher order integrations over $\Sigma$. 
In the limit $g\to 0$ (\ref{uno.1}) reduces to (\ref{uno.4}) which is
a topological theory. Making the perturbative expansion in $g$ around 
the topological theory, $\Delta(\Sigma)$ in (\ref{cin.13}) is found
to be a diffeomorphism invariant quantity, i.e. 
$\Delta (\Sigma )$ depends only on the knotting properties of $\Sigma$ 
and does not depend on the linking properties of $\Sigma$ 
(at least when our ambient space-time is $S^4$). 
If the
dynamical surface $\Sigma$ 
created by the quantum fluctuations of the true QCD vacuum is the unknotted 
surface $\Sigma_0$, one should find that 
\be 
\Sigma =\Sigma_0\rightarrow\Delta (\Sigma_0)=1\quad .
\label{cin.16}
\ee 
The explicit calculation of $\Delta (\Sigma )$ is an open 
problem, but for the purpose of showing the area law behaviour
its knowledge is not essential.  

Consider now the integral in the exponent of (\ref{cin.13}), 
\ba 
&& \oint_C\int_{\Sigma}<A*B> = 
<\oint_C A\int_{\Sigma}d\sigma^{\mu\nu}(x)(*B(x))_{\mu\nu} >\\
\nonumber 
&& =<\oint_C A\int_{\Sigma}(*d\sigma )^{\mu\nu}(x) B(x)_{\mu\nu} > =
\oint_C dy^{\lambda}\int_{\Sigma}(*d\sigma)^{\mu\nu}(x) <A(y)_{\lambda}
B(x))_{\mu\nu} >\ ,
\label{cin.29}
\ea 
where 
$*d\sigma (x)$ is the infinitesimal surface element of the plane $\Sigma_x^*$ 
dual to $\Sigma$ at the point $x\in\Sigma$. We may rewrite 
\be 
\oint_C dy^{\lambda}(*d\sigma)^{\mu\nu}(x) <A(y)_{\lambda}B(x))_{\mu\nu}>=
\oint_C dy^{\lambda}\int_{\Sigma^*_x}\omega_{\Sigma}^{\mu\nu} 
<A_{\lambda}B_{\mu\nu} > \quad ,
\label{cin.30}
\ee 
recalling that $\omega_{\Sigma}$ is locally given by a bump function with 
support on $\Sigma$. Eq. (\ref{cin.30}) is by definition the linking number 
between the curve $C$ and the dual plan $\Sigma^*_x$ in $x$ to $\Sigma$.  
Indeed the linking $\Link(C,\Sigma )$, whith arbitrary  $C$ and $\Sigma$, is 
defined by \cite{horo} 
\be 
\Link(C,\Sigma )=
\frac 1 {8\pi^2}\oint_{C}dx^{\alpha}
\int_{\Sigma}d\sigma^{\beta\gamma}(y)\epsilon_{\alpha\beta\gamma\delta}
{(x-y)^{\delta}\over |x-y|^4} 
=4\oint_{C}dx^{\alpha}
\int_{\Sigma}d\sigma^{\beta\gamma}(y)
<A_{\alpha}^a(x)B_{\beta\gamma}^a(y)>  \quad .
\label{cin.12}   
\ee  
In our case, by construction, $\Link(C,\Sigma^*_x)\neq 0$. The residual 
integration over $\Sigma$ in (\ref{cin.29}) spans all the dual $\Sigma^*$ 
to $\Sigma$, yielding a contribution proportional to the area of $\Sigma$, 
\ba  
<W_t(C)>_{conn} &\sim & \exp \{ -g^2{(N^2-1)\over 16N}
\int_{\Sigma} \Link(C,\Sigma^*_x)\} \quad , \nonumber \\ 
\int_{\Sigma} \Link (C,\Sigma^*_x) &\sim & A(\Sigma ) \quad . 
\label{cin.17}
\ea 
We may get a better understanding of (\ref{cin.17}) considering a lattice 
regularization of $\Sigma$, i.e. $\Sigma\to\Sigma_{PL}$. In this case $C$ 
runs over the links of the lattice, while $\Sigma_x^*$ corresponds to 
an element of the dual lattice through the plaquette centered at $x$. 
$\int_{\Sigma_{PL}}\Link(C,\Sigma^*_x)$ is an integer which 
counts the number $N_v$ of vertices 
of the dual lattice on $\Sigma_{PL}$ or equivalently the number of plaquettes 
$N_P$  of $\Sigma_{PL}$. We may then write
\be 
\int_{\Sigma_{PL}}\Link(C,\Sigma^*_x)= N_P={A(\Sigma_{PL})\over a^2}\quad ,
\label{cin.18} 
\ee  
where $A(\Sigma_{PL})$ is the minimal area 
bounded by the quark loop $\C$ in the fundamental representation and $a$ is 
the lattice spacing. 
Passing to the continuum limit  $a\to 0$, 
$N_P =A(\Sigma_{PL})/a^2$ becomes $A(\Sigma)/l^2$ with $l$ a typical
scale in QCD which may be choosen as $\Lambda_{QCD}^{-1}$.  
Therefore one may rewrite (\ref{cin.17}) as 
\be 
<W_t(\C)>_{conn}\sim \exp [-\sigma(\Lambda_{QCD}) A(\Sigma)]\quad ,
\label{cin.19}
\ee 
where $\sigma(\Lambda_{QCD})$ is the string tension defined by 
\be 
\sigma(\Lambda_{QCD}) \equiv g^2_R ({N^2-1\over 16N})
\Lambda^2_{QCD}\quad ,
\label{cin.19bis} 
\ee
with $\Lambda_{QCD}\equiv 1/l$.
Here we have replaced, as required by the renormalization of the theory, 
the bare coupling constant $g$ with the running coupling constant $g_R$ 
at some QCD energy scale $\Lambda_{QCD}$. Of course we may change the physical 
scale, but it is well known that this corresponds to a different choice of 
the renormalization scheme, the difference being only a finite 
renormalization, leading to the same physical results.

Existing data corresponding to energy scales between $4$ and 
$100(Gev)^2$ can 
be fitted with $\Lambda_{QCD}\sim 0.5$~$Gev$, corresponding to 
$[g^2_R(-p^2, \Lambda_{QCD})/4\pi ]$ between $0.2$ and $0.4$ for  
$ N_f=4$. Inserting these data in (\ref{cin.19bis}) 
we find that a rough estimate of the physical value of the string tension
can be given by 
\be 
\sqrt\sigma (\Lambda_{QCD}\sim 0.5Gev )\sim 316-440 Mev \quad ,
\label{cin.20}
\ee 
which is of the same order of magnitude of the experimental value.

Before concluding this chapter let us comment on the result we have found. 
It might seem puzzling that the area law behaviour comes out of a perturbative 
calculation. In reality our perturbative expansion is an expansion around 
the gauge fixed topological BF theory whose vacuum contains instantons. 
It is thus resonable to imagine that the non perturbative information comes out 
of the non-trivial vacuum structure of the quantum BF theory. 
Furthermore we remark 
that the computations of this section could also have been done along the lines 
of the chapter three, given the similarity between $M(\Sigma ,C)$
and $M^*(\Sigma ,C)$. As a consequence the invariant 
$Q(\Sigma ,\Sigma^{\prime})$ in (\ref{qua.30}) 
is replaced by $\Link(\Sigma^* ,C)$, 
in agreement with (\ref{cin.17}) and abelian dominance translates into 
(\ref{cin.16}), 
i.e. lack of interesting knotting properties of $\Sigma$.
 
\section{Conclusions}
\setcounter{equation}{0}
In this paper we have shown that in the framework of the BF formulation of 
Yang-Mills theory,
't Hooft views on confinement can be explicitly realized in the continuum 
formulation.
This has been possible because a non-abelian color magnetic operator has been
identified quite naturally in terms of a non local observable  (\ref{tre.2})
constructed out of 
the microscopic fields $A, B$. Then we have computed the average
of this operator in terms of a saddle point approximation having used the 
abelian
projection gauge. The implementation of this gauge has consisted in performing
the computation in terms of the large scale effective lagrangian obtained 
neglecting massive charged degrees of freedom. This point clearly awaits
further clarification and an estimate of the contribution of the discarded 
sector
remains to be evaluated. 
However, we do not believe that the perimeter law behaviour
in (\ref{qua.35}),
coming from the 
evaluation
of the classical action, can be 
disrupted
by quantum corrections.

In any case the ratio of two operators of the type (\ref{tre.2}) 
with
different paths $C, C^\prime$ still exhibits the perimeter law since any 
possible
divergence comes from shrinking to zero the size of the magnetic flux tube
(ANO string).  

In the framework we have investigated, the perimeter law 
stems from the breaking of the BF topological
theory (\ref{uno.4}) due to the $g^2\tr(B\wedge *B)/4$ term in 
(\ref{uno.1}). 
In \cite{cm} the expectation value of $M(C,\Sigma)$ was computed 
in the pure
BF theory looking for two-knot invariants i.e. for four dimensional 
topological
invariants of knotting of surfaces. In that computation a classical 
action 
piece like (\ref{qua.26}) did not appear because of the missing 
$\tr(B\wedge *B)$
term (which allows for gaussian integration) and only functions 
of the four 
dimensional topological invariant (\ref{qua.27}) were obtained. 
It is thus
reasonable that in the BF Yang-Mills appears a homotopy 
invariant of the
curve $C$ as $\sLink(C)$ which represent the effect of the winding 
of $C$ around
the monopole flux tube (which we can think closes at infinity). 
The curve  $C$ in (\ref{tre.2}) is connected to the holonomy of 
the gauge 
connection and contains the disks $S^2_\epsilon(x_{(i)})\subset\Sigma$.
These disks become the 
observable Dirac surfaces of  \cite{thooft1} and are responsible for the 
perimeter law. 
An argument similar has been developed in \cite{amati} for a BF Yang-Mills
theory in two dimensions where the two form $B$ becomes a scalar.

Let us now remark on some properties of (\ref{qua.35}) which agree with 
common 
wisdom: the appearence of $g_R$ has been confirmed in  
lattice simulations
of  $SU(N)$ QCD \cite{shiba}. The ``photons'' $Q_i$ interacting with a 
plasma of 
condensed monopoles reproduce the same $SU(N)$ running coupling 
constant 
required by the renormalization group. From (\ref{qua.26}) 
we have a perimeter
law behaviour for $m\neq 0$ and finite $N$ as expected for the confining
phase of QCD in the absence of massless particles.

As far as the average of the Wilson loop is concerned,
(\ref{cin.17}) is the basic result. 
Indeed it implies that the area
law for the Wilson  loop average
is a  consequence of the response of the QCD vacuum to arbitrary path 
deformations 
of the loop $\C$. This property is evident in our formalism, where 
the loop 
$\C=\partial\Sigma $ may be understood as the deformation of the 
loop $C$. The existence of $\Link(\Sigma^*,C)$ 
means that 
the expectation value $<W_R(\C)>$ is not invariant, as reasonably 
expected, 
under the change of the path-prescription $C$ in the non-abelian 
Stokes theorem given by (\ref{cin.1}).
In other words, (\ref{cin.17}) is a non-perturbative Ward identity 
associated 
to the path deformation of the static quark loop $\C$.
It is well-known that while the fundamental representation of the
gauge group leads to confinement, the adjoint representation does not.
This does not show up in (\ref{cin.17}) due to presence of the coupling
$g$ which should be absorbed by the presence of a
mass gap (from monopole condensation) in the propagator for
the $A$ and $B$ field.

Finally, observe that the results obtained in this paper are 
due to the non-abelian 
nature of the model. If we had started with a pure abelian 
BF-YM theory, 
the operator $M$ would have been defined without the {\rm Hol} 
terms 
and, therefore, without the possibility to have a 
non-trivial 
monopole equation (\ref{qua.19}). In the abelian case an 
additional 
Higgs field is needed 
in order to have a non-trivial magnetic variable $M$.

\vskip 1.5cm
\leftline{\bf\large Acknowledgments}
\vskip .5cm
The authors want to thank the groups of theoretical physics at the 
University
of Utrecht and at The Niels Bohr Institute for hospitality during 
part of this work and C.~Becchi, M.~Bianchi, A.~Cattaneo, P.~Cotta-Ramusino, 
E.~D'Emilio, G.~'t Hooft, M.~Mintchev  
and G.C.~Rossi for very stimulating conversations.
M. Martellini wants to thank J.~ Ambjorn, J.~Fr\" ohlich, 
J.~P.~Greensite and Y.~Makeenko for many 
long and interesting conversations.
M.Z. was partially supported by Ministero dell'Universit\`a e della Ricerca 
Scientifica e Tecnologica and by 
E.E.C. Grants CHRX-CT92-0035 and SC1$^*$-CI92-0789.

\vfill 
\eject 

\end{document}